\documentstyle[preprint,amssymb,aps,tighten,epsf,floats]{revtex}
\newcommand{\He}{^3\mbox{He}}
\newcommand{\nn}{\nonumber\\}
\newcommand{\ggama}{{\scriptstyle\Gamma}}

\newcommand{\la}{\langle}
\newcommand{\ra}{\rangle}

\newcommand{\bPsi}{\bar{\Psi}}

\newcommand{\ben}{\begin{displaymath}}
\newcommand{\een}{\end{displaymath}}
\newcommand{\be}{\begin{equation}}
\newcommand{\ee}{\end{equation}}
\newcommand{\bea}{\begin{eqnarray}}
\newcommand{\eea}{\end{eqnarray}}

\newcommand{\he}{\hat{e}}

\newcommand{\htt}{\hat{t}}

\newcommand{\bphi}{\bar{\phi}}

\newcommand{\NN}{$N\!N$}

\newcommand{\piNN}{$\pi N\!N$}

\newcommand{\NNN}{$N\!N\!N$}

\newcommand{\tU}{\tilde{U}}
\newcommand{\tX}{\tilde{X}}
\newcommand{\tY}{\tilde{Y}}
\newcommand{\tZ}{\tilde{Z}}
\newcommand{\tT}{\tilde{T}}
\newcommand{\mbf}[1]{\mbox{\boldmath {#1}}}

\newcommand{\bc}{\begin{center}}
\newcommand{\ec}{\end{center}}

\newcommand{\eqn}[1]{\label{#1}}
\newcommand{\eq}[1]{Eq.~(\ref{#1})}
\newcommand{\eqs}[1]{Eqs.~(\ref{#1})}
\newcommand{\fign}[1]{\label{#1}}
\newcommand{\fig}[1]{Fig.~\ref{#1}}

\begin{document}
\draft
\title{\bf Gauging of equations method.\\ 
II. Electromagnetic currents of three identical particles}
\author{A. N. Kvinikhidze\footnote{On leave from Mathematical Institute of
Georgian Academy of Sciences, Tbilisi, Georgia.} and B. Blankleider}
\address{Department of Physics, The Flinders University of South Australia,
Bedford Park, SA 5042, Australia}
\date{\today}
\maketitle
\begin{abstract}
  The gauging of equations method, introduced in the preceding paper, is applied
  to the four-dimensional integral equations describing the strong interactions
  of three identical relativistic particles. In this way we obtain gauge
  invariant expressions for all possible electromagnetic transition currents of
  the identical three-particle system. In the three-nucleon system with no
  isospin violation, for example, our expressions describe the electromagnetic
  form factors of $^3$H, $pd\rightarrow pd\gamma$, $\gamma
  ^3\mbox{He}\rightarrow pd$, $\gamma \He\rightarrow ppn$, etc. A feature of our
  approach is that gauge invariance is achieved through the coupling of the
  photon to all possible places in the (nonperturbative) strong interaction
  model.  Moreover, once the proper identical particle symmetry is incorporated
  into the integral equations describing the strong interactions, the gauging
  procedure automatically provides electromagnetic transition currents with the
  proper symmetry. In this way the gauging of equations method results in a
  unified description of strong and electromagnetic interaction of strongly
  interacting systems.
\end{abstract}


\section{Introduction}

In the preceding paper \cite{I} (referred to as I in the following), we have
introduced the gauging of equations method as a means of incorporating an
external electromagnetic field into descriptions of quarks or hadrons whose
strong interactions are described nonperturbatively by integral equations. The
feature of this method is that it couples an external photon to all possible
places in the strong interaction model despite its nonperturbative nature. Gauge
invariance in our approach is therefore implemented in the way prescribed by
quantum field theory (QFT). In I we demonstrated the gauging procedure on the
example of three distinguishable particles. In this paper we would like to
demonstrate the same method as applied to the case of indistinguishable
particles where the strong interaction equations have the added complexity of
identical particle symmetry.

As in I, the discussion here is restricted to the case where the three strongly
interacting particles have no coupling to two-body channels. Thus we have in
mind identical particle systems like three quarks $qqq$ or three nucleons \NNN.
This is not an essential restriction, and indeed we have recently applied the
gauging of equations method to the \piNN\ system where coupling to the \NN\ 
channel is included \cite{gpinn}. However, the purpose of this paper (together
with I) is to present the basic details of the gauging of equations method, and
as such, coupling to two-body channels presents an unnecessary complication.  We
note that the main results of this work have previously been summarised in
conference proceedings \cite{previous}.

In dealing with identical particles, one is faced with the problem of
incorporating the proper particle-exchange symmetry into the equations
describing both their strong and electromagnetic interactions. In Quantum
Mechanics the standard procedure is to explicitly symmetrise (or antisymmetrise)
the corresponding distinguishable particle equations. However, such a procedure
is not justified within a field theoretic approach. As the basis of our approach
here is QFT, we present a derivation of the strong interaction equations for
three identical particles that is consistent with QFT, and that is therefore
very different from the derivations found in standard texts on the quantum
mechanical three-body problem \cite{AT}.

Once the identical particle equations for the strong interactions are derived,
it is a feature of the gauging of equations method that it may be applied
directly to these equations, thereby automatically generating electromagnetic
transition currents with the proper symmetry. Thus the main effort in generating
practical expressions for the transition currents reduces down to a careful
choice for the identical particle strong interaction equations that are to be
gauged. Here we also show that for three identical particles an optimal choice
is provided by an equation of Alt-Grassberger-Sandhas (AGS) form \cite{AGS}, but
which is however different from the AGS equation used previously in the
literature for three-nucleon calculations.

\section{Gauging the three-particle Green function}

The gauging of equations method, introduced in I and used there to gauge the
equations of three distinguishable particles, does not change when the particles
are identical. Indeed, the main steps taken in I to derive the electromagnetic
transition currents can be repeated for identical particles, although it is
necessary to guarantee the proper identical particle symmetry of all
perturbative diagrams at each step of the derivation. In this respect, it should
be noted that we do not follow the common procedure of symmetrising or
antisymmetrizing the distinguishable particle results. Such a procedure is
strictly valid only within the context of second quantisation in quantum
mechanics, while here the theoretical framework is that of relativistic quantum
field theory. Instead we follow the standard rules of QFT for constructing 
Green functions for identical particles. Details of these rules as applied to
few-body integral equations have been given by us in Ref.\ \cite{KB4d}.

The strong interactions of three identical particles are described in quantum
field theory by the Green function $G$ defined by
\bea
\lefteqn{(2\pi)^4\delta^4(p'_1+p'_2+p'_3-p_1-p_2-p_3)G(p'_1p'_2p'_3;p_1p_2p_3)=
\int d^4y_1d^4y_2d^4y_3d^4x_1d^4x_2d^4x_3} \hspace{1.5cm}\nn
&&e^{i(p'_1y_1+p'_2y_2+p'_3y_3-p_1x_1-p_2x_2-p_3x_3)}\,
\la 0|T\Psi(y_1)\Psi(y_2)\Psi(y_3)
\bar{\Psi}(x_1)\bar{\Psi}(x_2)\bar{\Psi}(x_3)|0\ra. \eqn{G6pt}
\eea
Here $\Psi$ and $\bar{\Psi}$ are Heisenberg fields, $T$ is the time ordering
operator, and $|0\ra$ is the physical vacuum state. The interaction of this
three-particle system with an external electromagnetic field is then
described by the corresponding seven-point function $G^\mu$ defined by 
\bea
\lefteqn{G^\mu (k_1k_2k_3;p_1p_2p_3)=
\int d^4y_1d^4y_2d^4y_3d^4x_1d^4x_2d^4x_3}
\hspace{0cm}\nn
&&e^{i(k_1y_1+k_2y_2+k_3y_3-p_1x_1-p_2x_2-p_3x_3)}
\la 0|T\Psi(y_1) \Psi(y_2)\Psi(y_3)
\bar{\Psi}(x_1)\bar{\Psi}(x_2)\bar{\Psi}(x_3)J^\mu(0)|0\ra
\eqn{G7pt}
\eea
where $ J^\mu$ is the quantised electromagnetic current operator and $e_i$ is
the charge of the $i$-th particle. If the particles are isotopic doublets then
$e_i$ includes an isospin factor, e.g.\ for nucleons
$e_i=\frac{1}{2}[1+\tau_3^{(i)}]e_p$ where $\tau_3$ is the Pauli matrix for the
third component of isospin, and $e_p$ is the charge of the proton. The
Ward-Takahashi [WT] identity \cite{WT}, which provides an important constraint
on $G^\mu$, takes the same form as in the distinguishable particles case
\cite{I}
\bea 
\lefteqn{ q_\mu G^\mu (k_1k_2k_3;p_1p_2p_3)=
i[e_1G(k_1-q,k_2k_3;p_1p_2p_3)+e_2G(k_1,k_2-q,k_3;p_1p_2p_3)}
\hspace{3cm}\nn
&&+e_3G(k_1k_2,k_3-q;p_1p_2p_3)-G(k_1k_2k_3;p_1+q,p_2p_3)e_1\nn
&&-G(k_1k_2k_3;p_1,p_2+q,p_3)e_2-G(k_1k_2k_3;p_1p_2,p_3+q)e_3] ,
\hspace{1cm} \eqn{WGmu}
\eea
or in the shorthand notation introduced in Ref.\ \cite{I},
\be
q_\mu G^\mu (k_1k_2k_3;p_1p_2p_3)=i\sum_{i=1}^3
[e_iG(k_i-q;p_1p_2p_3)-G(k_1k_2k_3;p_i+q)e_i].
\ee

To be definite, we shall assume that our three identical particles are fermions
(for three identical bosons one can simply replace antisymmetric operations by
symmetric ones in the following discussion).  Then the field theoretic
expressions of \eqs{G6pt} and (\ref{G7pt}) automatically guarantee the proper
antisymmetry of the three-particle Green function $G$ and the seven-point
function $G^\mu$.  On the other hand, the free Green function $G_0$ defined by
\be
G_0=d_1d_2d_3    \eqn{G_0123}
\ee
where $d_i$ is the dressed propagator of particle $i$, is not antisymmetric in
its particle labels; thus, for identical fermions, $G_0$ is not equal to the
fully disconnected part of $G$ (which we shall denote by $G_d$).  Indeed, it can
be easily shown \cite{KB4d} that to obtain $G_d$, one needs only to
antisymmetrize $G_0$ according to the equation
\be
\sum_P G_0(1'2'3',123) = G_d(1'2'3',123)     \eqn{calG_0}
\ee
where the sum is over all permutations $P$ of either the initial or final state
particle labels, and is understood to include a factor $(-1)^P = +1$ or $-1$
depending on whether the permutation is even or odd, respectively. In
\eq{calG_0} we use a symbolic notation where integers represent the momenta and
all quantum numbers of the corresponding particles, with primes distinguishing
the final states. To specify permutation sums over just the initial state
momentum labels, we use the letter $R$ (right); similarly, $L$ (left) represents
sums over permutations of just the final state momentum labels.  The symbol $P$
will be used only when it makes no difference which sum $R$ or $L$, is
taken. Quantities antisymmetrized in one of these ways will be indicated by the
appropriate superscript. Thus, for example, if $A$ is a quantity depending on
three initial and three final particle labels, then
\bea
A^R(1'2'3',123)&=& A(1'2'3',123)-A(1'2'3',213)+A(1'2'3',231)- \ldots \\
A^L(1'2'3',123)&=& A(1'2'3',123)-A(2'1'3',123)+A(2'3'1',123)- \ldots
\eea
with similar expressions holding for quantities having any number of identical
legs.  In general we write:
\bea
A^R &\equiv& \sum_R  A\\ A^L &\equiv& \sum_L  A\\
A^P &\equiv& \sum_P  A = A^R = A^L.
\eea

Defining the kernel $K$ to be the set of all possible three-particle irreducible
Feynman diagrams for the $3\rightarrow 3$ process, we may write the Green
function $G$ as \cite{KB4d}
\be
G = G_0^P + \frac{1}{3!} G_0 K G       \eqn{G_as}
\ee
where the $1/3!$ factor reflects the fact that both $G$ and $K$ are fully
antisymmetric in their particle labels. We write the disconnected part of $K$,
indicated by subscript $d$, in terms of the identical particle two-body
potential $v$:
\be
K_d(1'2'3',123)=\sum_{L_cR_c} v(2'3',23) d^{-1}(1)\delta(1',1)
\ee
where $\delta(1',1)$ represents the momentum conserving Dirac $\delta$ function
$(2\pi)^4\delta^4(p_1'-p_1)$, while $L_c$ and $R_c$ indicate that the sum
is taken over cyclic permutations of the left labels $(1'2'3')$ and right labels
$(123)$, respectively (note that the sum is restricted to cyclic permutations
because the potential $v$ is already antisymmetric in its labels \cite{KB4d}).

Defining
\be
V_i(1'2'3,123) = v(j'k',jk) d^{-1}(i)\delta(i',i)      \eqn{V_i_sym}
\ee
where $(ijk)$ is a cyclic permutation of $(123)$, we have that
\be
K_d = \sum_{P_c} (V_1+V_2+V_3),
\ee
where it makes no difference over which labels, left or right, the cyclic
permutations are taken. Unlike the $V_i$ of the distinguishable particle case
[see Eq.(61) of I], the one here consists of a two-body potential $v$ that is
antisymmetric under the interchange of its initial or final state labels.
Denoting the connected part of the kernel by $K_c$, we define the $3\rightarrow
3$ potential $V$ by
\be
V=\frac{1}{2}(V_1+V_2+V_3) + \frac{1}{6}K_c.   \eqn{V_sym}
\ee
Although $V$ is not fully antisymmetric, it does have the useful symmetry
property
\be
P_{ij} V P_{ij} = V    \eqn{V-ij-symm}
\ee
where $P_{ij}$ is the operator that exchanges the $i$'th and $j$'th momentum,
spin, and isospin labels. Since
\be
K=\sum_P  V,
\ee
\eq{G_as} can be written as
\be
G = G_0^P + G_0 V G .      \eqn{Gas}
\ee
Formally, \eq{Gas} differs from the equivalent relation for distinguishable
particles [Eq.(59) of I] only in the explicit antisymmetrization of the
inhomogeneous term. We therefore proceed as for the distinguishable particle
case and gauge \eq{Gas} directly:
\be
G^\mu =G_0^{P\mu} + (G_0V)^\mu G + G_0VG^\mu .  \eqn{duck}
\ee
Before solving this equation for $G^\mu$ it is useful to note that
\be
G_0^{P\mu}=G_0^{\mu P}=G_0^PG_0^{-1}G_0^\mu .
\ee
Indeed the combination $G_0^PG_0^{-1}$ plays the role of the antisymmetrization
operator since
\be
[G_0^PG_0^{-1}](1'2'3,123)=\sum_P \delta(1',1) \delta(2',2) .
\eqn{P}
\ee
Note that there is no factor $\delta(3',3)$ on the right side of \eq{P} because
an overall momentum conservation delta-function has been removed from our
expressions. We can therefore write \eq{duck} as
\be G^\mu = (1-G_0V)^{-1} \left[ G_0^PG_0^{-1}G_0^\mu +
G_0^PG_0^{-1}\frac{1}{6}(G_0V)^\mu G \right]
\ee
where the inclusion of $G_0^PG_0^{-1}$ in the last term is compensated exactly
by $1/6$ since $V$ satisfies the symmetry property of \eq{V-ij-symm} and $G$
is already fully antisymmetric. From \eq{Gas} it follows that
\be (1-G_0V)^{-1}G_0^P = G,
\ee
and using this we obtain
\bea G^\mu &=& G G_0^{-1}\left[ G_0^\mu + \frac{1}{6} (G_0^\mu V G + G_0 V^\mu
G) \right] \\
&=& \frac{1}{6} G \left[ G_0^{-1}G_0^\mu G_0^{-1}(6G_0+G_0 V G )
+ V^\mu G \right].
\eea
In the last equation, we may replace the $6G_0$ by $G_0^P$ because of the
antisymmetry of the $G$ outside the square bracket, and using \eq{Gas} we
finally get that
\be G^\mu = G\Gamma^\mu G \eqn{yahoo}
\ee
where
\be 
\Gamma^\mu = \frac{1}{6}\left( G_0^{-1}G_0^\mu G_0^{-1}+ V^\mu \right)
\eqn{Gamma^mu_sym}
\ee
is the electromagnetic vertex function for three identical particles.
The extra factor of $1/6$ compared with the result for distinguishable
particles reflects the fact that here $\Gamma^\mu$ is to be sandwiched between
fully antisymmetric functions. Neglecting three-body forces, \eq{V_sym} implies
that
\be
V^\mu = \frac{1}{2}\left( V_1^\mu + V_2^\mu + V_3^\mu \right)
\ee
with an extra factor of $1/2$ compared with the distinguishable particle case.

Writing \eq{V_i_sym} in the shorthand notation
\be
V_i=v_i d_i^{-1}      \eqn{V_i}
\ee
where $v_i$ denotes $v(j'k',jk)$, we may gauge this equation to obtain
\be
V_i^\mu=v^\mu d_i^{-1}-v_i\ggama_i^\mu
\ee
where $\ggama_i^\mu$ is the one-particle electromagnetic vertex function defined
by the equation
\be
d_i^\mu=d_i\ggama_i^\mu d_i,
\ee
and where we have used the fact that $(d_i^{-1})^\mu=-\ggama_i^\mu$ \cite{I}.
Similarly gauging \eq{G_0123} we find that
\be
G_0^{-1}G_0^\mu G_0^{-1} = \sum_{i=1}^{3}  \ggama_i^\mu D_{0i}^{-1}
\ee
where
\be
D_{0i}=d_jd_k.
\ee
Using these results in \eq{Gamma^mu_sym} we can express the electromagnetic
vertex function as
\be
\Gamma^\mu=\frac{1}{6}\sum_{i=1}^3 \left(\ggama_i^\mu D_{0i}^{-1} +\frac{1}{2}
v_i^\mu d_i^{-1} -\frac{1}{2} v_i \ggama_i^\mu \right) . \eqn{howdy}
\ee

All electromagnetic transition currents of three identical particles can be
obtained from \eq{yahoo} by taking appropriate residues at two- and three-body
bound-state poles of $G$. If $G$ admits a three-body bound state, it can be
shown that
\be
G(p'_1p'_2p'_3;p_1p_2p_3) \sim i \frac{\Psi_P(p'_1p'_2p'_3)
\bar{\Psi}_P(p_1p_2p_3)}{P^2-M^2} \hspace{1cm}\mbox{as} 
\hspace{5mm} P^2\rightarrow M^2                     \eqn{Gpole}
\ee
where $P$ is the total momentum, $M$ is the bound-state mass, and $\Psi_P$ is
the three-particle bound-state wave function defined by
\bea
\lefteqn{(2\pi)^4 \delta^4(P-p_1-p_2-p_3)\Psi_P(p_1p_2p_3)  = }\hspace{2cm} \nn
&& \int d^4x_1d^4x_2d^4x_3 e^{i(p_1x_1+p_2x_2+p_3x_3)}
\la 0|T\Psi(x_1)\Psi(x_2)\Psi(x_3)|P\ra .      \eqn{Psi_P}
\eea
Here $|P\ra$ is the eigenstate of the full Hamiltonian corresponding to the
three-particle bound state with momentum $P^\mu$. 

The three-body bound-state current $j^\mu$ is found by taking
left- and right-residues of $G^\mu$ at the three-body bound-state poles.
By exposing such poles in the field theoretic expression of \eq{G7pt} one finds
that
\be
j^\mu = \la J^\mu \ra \equiv \la K|J^\mu(0)|P\ra    \eqn{j^mu_exp}
\ee
where $K^2 = P^2 = M^2$. To find $j^\mu$ for a particular model, one can
alternatively use \eq{yahoo} to take residues at the three-body bound-state
poles. In this way one finds that
\be
j^\mu  =\bar{\Psi}_{K}\Gamma ^\mu \Psi_P .         \eqn{j^mu1_identical}
\ee
Although this expression is formally identical to the bound-state current for
distinguishable particles [Eq.(80) of I], here the vertex function is given by
\eq{howdy}, and the wave function is that for identical particles.

The normalisation condition for the wave function in the case of identical
particles follows from the fact that
\be
G\left(G_0^{-1}-V\right)G = G G_0^{-1}G_0^P = 6G.
\ee
After taking residues at the three-body bound-state poles of $G$ one finds that
\be 
i \bar{\Psi}_P \frac{\partial}{\partial P^2}\left(G_0^{-1}-V\right)\Psi_P =
6. \eqn{wave_norm_id} 
\ee
Note that this result differs from the one for distinguishable particles where
unity appears on the right hand side [Eq.(82) of I]. This is a consequence of
our convention where the same expression, \eq{Psi_P}, is used to define the
bound-state wave function for both identical and distinguishable particles
(in the latter case, however, the fields obtain particle labels).

We can repeat the above procedure and take residues of \eq{yahoo} at two-body
bound-state poles of $G$, thereby obtaining electromagnetic transition currents
involving two-body bound states. This procedure was carried out in detail for
the distinguishable particle case in I. However, as discussed in I, the
expressions obtained in this way explicitly involve potentials and gauged
potentials, and consequently may not be very convenient for practical
calculations. Here we shall therefore forego any further discussion of this
procedure, and instead go on to an alternative approach based on the
Alt-Grassberger-Sandhas (AGS) equations \cite{AGS} which lead, after gauging, to
electromagnetic transition currents expressed in terms of $t$-matrices and
gauged $t$-matrices.

\section{AGS amplitudes for identical particles}

The AGS equations have long provided a practical way to describe the scattering
of three particles in Quantum Mechanics. Not only do they lead (after one
iteration) to equations with a connected kernel, but they also have the feature
of having the two-body inputs in terms of $t$-matrices rather than potentials.
As we would also like to have these advantages in the case of three relativistic
particles, we shall utilise four-dimensional versions of the AGS equations,
which for distinguishable particles were given by Eqs.(118) of I.

Our goal here is to extend the discussion of Sec.\ III~C of I to the case of
three identical particles. That is, we would like to express the electromagnetic
transition currents of all possible processes involving three identical
particles in terms of AGS amplitudes and gauged AGS amplitudes. Although the
handling of identical particles in the AGS formulation is well documented
\cite{AT}, as far as we know all previous discussions do this by
antisymmetrizing the distinguishable particle case. As stated previously, such a
procedure is inappropriate for the field theoretic approach undertaken here.  In
this section we shall therefore define AGS amplitudes for identical particles
and relate them to the $3\rightarrow 3$ Green function $G$ in a way that is
consistent with field theory.  Moreover, we do this with the view of gauging our
final expressions, a task left to the following sections.

Our starting point here shall be the Green function for three identical
particles as given by \eq{Gas}. The natural way to introduce the AGS operators
$U_{ij}$ for three identical particles is via the distinguishable particle
case. Recalling that \eq{Gas} differs from the one for distinguishable particles
in that the inhomogeneous term $G_0^P$ is explicitly antisymmetrized, we are led
to introduce a new three-particle Green function $G^D$ defined by the equation
\be
G^D=G_0+G_0VG^D \eqn{G^D}
\ee
where the inhomogeneous term $G_0$ is not antisymmetrized. By its structure,
\eq{G^D} looks like the equation for the Green function of three distinguishable
particles [Eq.(59) of I] and therefore allows us to define the AGS operators in
the standard way.  Nevertheless, $G^D$ should not be identified with the
distinguishable particle Green function as the $V$ in \eq{G^D} is defined in
terms of antisymmetric potentials $v$ [see \eqs{V_i_sym} and (\ref{V_sym})]
while the $V$ for distinguishable particles is defined in terms of two-body
potentials which are not antisymmetric.  The fully antisymmetric Green function
$G$ can be obtained from $G^D$ simply by antisymmetrizing:
\be
G=G^{DL}=G^{DR}=G^{DP}.
\ee

Neglecting the three-body force $V_c$, we now proceed by analogy with the 
distinguishable particles case and define the AGS operators $U_{ij}$ through the
equation
\be
G^D= G_i\delta_{ij}+G_i U_{ij} G_j   \eqn{G^D=U_ij}
\ee
where $G_i$ satisfies the equation
\be
G_i=G_0+\frac{1}{2}G_0V_iG_i     \eqn{G_i}
\ee
(note that the inhomogeneous term in the last equation is not antisymmetrized,
in contrast to the $G_i$ used for identical particles in Ref.\ \cite{KB4d}). The
factor of $1/2$ in \eq{G_i} originates from \eq{V_sym} where it is clear that
$1/2 V_i$ is the disconnected potential to be identified with the
distinguishable particle case. Taking this into account, the AGS equations for
the operators $U_{ij}$ become
\be
U_{ij}=G_0^{-1}\bar{\delta}_{ij}+\frac{1}{2}\sum_{k=1}^{3}\bar{\delta}_{ik}T_k
G_0U_{kj} \hspace{5mm}; \hspace{5mm} U_{ij}=G_0^{-1}\bar{\delta}_{ij}+
\frac{1}{2}\sum_{k=1}^{3}U_{ik}G_0T_k\bar{\delta}_{kj} \eqn{AGS_eq_sym}
\ee
where the $T_i$ satisfy the equation
\be
T_i = V_i + \frac{1}{2}V_i G_0 T_i         \eqn{T_i}
\ee
and are given in terms of the two-body $t$-matrices $t_i$ by
\be
T_i=t_i d_i^{-1}.
\ee
Note that $t_i$ is shorthand for $t(j'k',jk)$ and is fully antisymmetric under
the interchange of its initial or final particle labels.

As discussed in I, for gauging purposes it is preferable to work in terms of AGS
Green functions
\be
\tU_{ij} = G_0 U_{ij} G_0 
\ee
which now satisfy the equations
\be
\tU_{ij}=G_0\bar{\delta}_{ij}+\frac{1}{2}\sum_{k=1}^{3}\bar{\delta}_{ik}G_0T_k
\tU_{kj} \hspace{5mm}; \hspace{5mm} \tU_{ij}=G_0\bar{\delta}_{ij}+
\frac{1}{2}\sum_{k=1}^{3}\tU_{ik}T_kG_0\bar{\delta}_{kj}. \eqn{tildeU_sym}
\ee
By using the above equations one can show that
\be
G^D=G_0+\frac{1}{2}\sum_{i}G_0T_iG_0 + \frac{1}{4}\sum_{i,k}G_0
T_i\tU_{ik}T_k G_0 .
\ee
In this way we obtain that
\bea
G&=&
G_0^P+\frac{1}{2}\sum_{i}G_0T_iG_0^P +
\frac{1}{4}\sum_{i,k}G_0T_i\tU_{ik}T_k G_0^P \eqn{UR} \\
&=&
G_0^P+\frac{1}{2}\sum_{i}G_0^PT_iG_0 +
\frac{1}{4}\sum_{i,k}G_0^PT_i\tU_{ik}T_k G_0.
\eea
By taking appropriate residues, either of these relations allow us to obtain the
scattering amplitudes for all possible processes in the system of three
identical particles. To be specific, let us choose \eq{UR} to discuss the taking
of residues.  As only the last term in \eq{UR} is connected, only this
term need be considered for the purposes of extracting physical three-particle
amplitudes. We therefore define
\be
G_c = \frac{1}{4}\sum_{i,k}G_0T_i\tU_{ik}T_k G_0^P .  \eqn{GU_ik}
\ee
By writing out this sum explicitly and making use of the symmetry properties of
$\tU_{ik}$ discussed in Appendix A, it is possible to rewrite \eq{GU_ik} in
terms of just one AGS-like Green function $\tU$:
\be
G_c = \sum_{L_cR_c} G_0 T_1\tU T_1 G_0      \eqn{GU_short}
\ee
A detailed derivation of \eq{GU_short} is presented in Appendix C. The full
Green function can then be written as
\be
G = G_0^P + \sum_{L_cR_c}\left(G_0 T_1 G_0 + G_0 T_1\tU T_1 G_0\right).
\eqn{Gid}
\ee
As discussed in Appendix C, there is a variety of ways to choose $\tU$
without affecting the value of $G$. One form that has previously been used
in three-nucleon calculations \cite{Glockle} is
\be
\tU = \frac{1}{2}\tX    \eqn{badU}
\ee
with $\tX$ obeying the equation [see \eq{X}]
\be
\tX = G_0{\cal P} + \frac{1}{2}{\cal P}G_0T_1\tX
\eqn{tX}
\ee
where ${\cal P} = P_{12}P_{31}+P_{31}P_{12}$ is the sum of two successive cyclic
permutations. The choice for $\tU$ specified by \eq{badU} is unsatisfactory for
our purposes since the presence of a sum of two permutations in both the
inhomogeneous term and kernel of \eq{tX} makes the gauging of this expression
particularly cumbersome. Fortunately there is another way to choose $\tU$ that
avoids these difficulties. As shown in Appendix C, we can take
\be
\tU = -\tZ P_{12}           \eqn{UZ}
\ee
where the AGS-like Green function $\tZ$ obeys the equation
\be
\tZ = G_0 - G_0 P_{12}T_1\tZ                           \eqn{tZ}
\ee
with no permutation sums involved. This is the form for $\tU$ that we
shall use in the next section for the purposes of gauging.
By displaying all momentum variables in \eq{GU_short} it is easy to see that
the connected part of the Green function can be written directly in terms of
$\tZ$ as
\be
G_c = \sum_{L_cR_c}G_0 T_1 \tZ T_2 G_0    .            \eqn{GUZ}
\ee  

\subsection{$Nd\rightarrow Nd$ amplitude}

For notational purposes, we shall refer to our three identical strongly
interacting fermions as a ``nucleons'' (\NNN) although the true identity of
these particles is arbitrary. Similarly we refer to a two-body bound state as a
``deuteron'' ($d$) and a three-body bound state as ``$^3$H''. This enables us
to write the various reactions that can take place between any three identical
particles in a familiar way.

Using this notation, we can obtain the amplitude for $Nd\rightarrow Nd$, by
follow the usual procedure of taking residues at the two-body bound-state poles
of $G$. Indeed for identical particles it can be shown that if quantum field
theory admits the existence of two-body bound states, then the Green function
$G(k_1k_2k_3;p_1p_2p_3)$ possesses poles with respect to any of the variables
$(k_i+k_j)^2$ or $(p_i+p_j)^2$. To be definite, consider the variables
$(k_2+k_3)^2$ and $(p_2+p_3)^2$ for the three-particle system.  In the vicinity
of the corresponding two-body poles, only the connected part of $G$ contributes
and we have that
\be
G_c(k_1k_2k_3;p_1p_2p_3)\sim
i\frac{\psi_{K_1}(k_2k_3)d(k_1)}{(k_2+k_3)^2-m^2}\,\, T_{dd}(k_1K_1;p_1P_1)\,\, 
i\frac{d(p_1)\bar{\psi}_{P_1}(p_2p_3)}{(p_2+p_3)^2-m^2} \eqn{1,1}
\ee
where $K_1=k_2+k_3$, $P_1=p_2+p_3$, $\psi_{K_1}$ is the deuteron wave function
[defined analogously to \eq{Psi_P}],
$m$ is the mass of the deuteron, and
$T_{dd}(k_1K_1;p_1P_1)$ is the physical scattering amplitude for
$N(p_1)+d(P_1)\rightarrow N(k_1)+d(K_1)$. The same scattering amplitude could be
picked out in the other channels, for example
\be
G_c(k_1k_2k_3;p_1p_2p_3)\sim
i\frac{\psi_{K_1}(k_2k_3)d(k_1)}{(k_2+k_3)^2-m^2}\,\, T_{dd}(k_1K_1;p_3P_3)\,\,
i\frac{d(p_3)\bar{\psi}_{P_3}(p_1p_2)}{(p_1+p_2)^2-m^2}
\ee
where $T_{dd}(k_1K_1;p_3P_3)$ depends on variables $p_3$, $P_3$ in just the same
way as $T_{dd}(k_1K_1;p_1P_1)$ depends on $p_1$, $P_1$. On the other hand, $G_c$
is given by \eq{GU_short} which when written out explicitly reads
\bea
\lefteqn{G_c(k_1k_2k_3;p_1p_2p_3) =
\sum_{L_cR_c}\int dk_2'\,dp_2'\,D_0(k_2k_3) t(k_2k_3;k'_2k'_3)}\hspace{5cm}\nn
&& 
\tU(k_1k'_2k'_3;p_1p'_2p'_3)t(p'_2p'_3;p_2p_3)D_0(p_2p_3) \eqn{gu2}
\eea
where $k'_2+k'_3=K_1$ and $p'_2+p'_3=P_1$.  In \eq{gu2} $D_0$ is the free
two-body propagator, and the poles at $(k_2+k_3)^2=m^2$ and $(p_2+p_3)^2=m^2$
are contained in the two-body $t$-matrices $t(k_2k_3;k'_2k'_3)$ and
$t(p'_2p'_3;p_2p_3)$, respectively. In particular,
\be
t(k_2k_3;k'_2k'_3)\sim i\frac{\phi_{K_1}(k_2k_3) \bar{\phi}_{K_1}(k'_2k'_3)}
{K_1^2-m^2}
\ee
where $\phi$ and $\bphi$ are two-body bound-state vertex functions defined by 
the equations
\be
\psi = G_0\phi\hspace{5mm};\hspace{5mm} \bar{\psi} = \bar{\phi}G_0. \eqn{phi}
\ee
Note that the two-body
$t$-matrix satisfies the integral equation
\be
t=v+\frac{1}{2}v D_0 t.     \eqn{t}
\ee
With particle $i$ as spectator, this equation when multiplied by $d_i^{-1}$
gives \eq{T_i}. The full two-body Green function $D$ is given by analogy 
with \eq{G_as} as
\be
D = D_0^P+\frac{1}{2}D_0 v D.      \eqn{D_as}
\ee
The normalisation condition for our two-body wave function is therefore given by
\be
i\bar{\psi}_P\frac{\partial}{\partial P^2}\left(D_0^{-1}-\frac{1}{2}
v\right)\psi_P=2,
\ee
a fact that follows from the same argument that led to \eq{wave_norm_id} but
adapted to the case of two identical particles.

Comparing \eq{1,1} with \eq{gu2} in the vicinity of the two poles gives the $Nd$
elastic scattering amplitude as
\be
T_{dd}(k_1K_1;p_1P_1)= \int dk_2'\,dp_2'\, d^{-1}(k_1)\bar{\phi}_{K_1}(k'_2k'_3)
\tU(k_1k'_2k'_3;p_1p'_2p'_3)
\phi_{P_1}(p'_2p'_3)d^{-1}(p_1)  .
\ee
This equation can be written symbolically as
\be
T_{dd} = d_1^{-1}\bar{\phi}_1 \tU\,\phi_1
d_{p_1}^{-1}                                     \eqn{Td}
\ee
or in terms of deuteron wave functions as
\be
T_{dd} = \bar{\psi}_1 U\,\psi_1.
\ee

\subsection{$Nd\rightarrow NNN$ amplitude}

The amplitude for the breakup reaction $Nd\rightarrow NNN$ is found by taking
the residue of the Green function $G$ at the initial state deuteron pole.
Choosing momentum variables as above, we may express the connected part
of the Green function $G_c$ in the vicinity of the initial bound-state pole as
\be
G_c(k_1k_2k_3;p_1p_2p_3)\sim
d(k_1) d(k_2) d(k_3)\, T_{0d}(k_1k_2k_3;p_1P_1)\,\, 
i\frac{d(p_1)\bar{\psi}_{P_1}(p_2p_3)}{(p_2+p_3)^2-m^2} \eqn{0,1}
\ee
where $(p_2+p_3)^2\rightarrow m^2$. This relation defines the breakup amplitude
$T_{0d}(k_1k_2k_3;p_1P_1)$. Comparing this with the behaviour of \eq{gu2} in
the vicinity of the same pole we deduce that
\bea
\lefteqn{T_{0d}(k_1k_2k_3;p_1P_1)= \sum_{L_c} \int dk_2'\,dp_2'}
\hspace{3cm}   \nn
&& d^{-1}(k_1)t(k_2k_3;k'_2k'_3)\tU(k_1k'_2k'_3;p_1p'_2p'_3)
\phi_{P_1}(p'_2p'_3)d^{-1}(p_1).   
\eea
Written symbolically this gives
\be
T_{0d} = \sum_{L_c} T_1 \tU \phi_1 d_1^{-1}
= \sum_{L_c} T_1 G_0 U \psi_1.    \eqn{T0d}
\ee
Using \eq{UZ} and \eq{tZ} we may eliminate $T_1$ from the last equation to
obtain
\be
T_{0d} = \sum_{L_c}(P_{12}ZP_{12}-G_0^{-1})\psi_1. \eqn{T0d_new}
\ee

\section{Gauging the identical particle AGS equations}

Having developed the necessary expressions describing three identical particles
in the purely strong interaction sector, we are now ready to carry out the
gauging procedure that will generate the coupling to an external electromagnetic
field. We follow the same procedure as used for distinguishable particles
in Sec.\ III~C of I.

\subsection{$Nd\rightarrow Nd$ transition current}

The $Nd\rightarrow Nd$ electromagnetic transition current $j_{dd}^\mu$
describes, for example, the process $Nd\rightarrow \gamma Nd$.  To obtain the
expression for $j_{dd}^\mu$ we write \eq{Td} as
\be
\tT_{dd} = \bar{\phi}_1 \tU\,\phi_1  \eqn{tTd}
\ee
where
\be
\tT_{dd} =d_1T_{dd}d_1.
\ee
Gauging \eq{tTd} gives
\be
\tilde{T}_{dd}^\mu=\bar{\phi}_1^\mu \tU\phi_1 
+ \bar{\phi}_1\tU\phi_1^\mu + \bar{\phi}_1\tU^\mu\phi_1 \eqn{tT^mu}.
\ee
The $Nd\rightarrow Nd$ electromagnetic transition current is then given by
\be
j_{dd}^\mu = d_1 ^{-1}\left(\bar{\phi}_1 ^\mu \tU\phi_1  
+ \bar{\phi}_1 \tU\phi_1 ^\mu + \bar{\phi}_1 \tU^\mu\phi_1  \right)d_1 ^{-1}.
\eqn{j_dd^mu}
\ee 
In \eq{tT^mu} $\bar{\phi}_1 ^\mu$ and $ \phi_1 ^\mu$ are the gauged two-body
bound-state vertex functions which follow from the solution of the two-body
problem for particles 2 and 3. Dropping the spectator particle label, the
bound-state equation for $\phi$ is
\be
\phi=\frac{1}{2}vD_0\phi.
\ee
Gauging this equation and using \eq{D_as} gives 
\be
\phi^\mu=\frac{1}{4}D^{-1}_0D[vD_0]^\mu\phi.
\ee
By using the fact that
\be
D=G_0^P+D_0 t D_0,
\ee
the previous equation can also be written as
\be
\phi^\mu=(1+\frac{1}{2}tD_0)(D_0^{-1}D_0^\mu D_0^{-1}+\frac{1}{2}v^\mu)\psi
-D_0^{-1}D_0^\mu D_0^{-1}\psi
\ee
which is the identical particle version of Eq.(38) of I.

To determine $j_{dd}^\mu$ all that is left is to specify a practical expression
for $\tU^\mu$. If we make the choice $\tU=-\tZ P_{12}$ as in \eq{UZ}, then
$\tU^\mu=-\tZ^\mu P_{12}$ and the problem reduces to that of gauging \eq{tZ}
in order to obtain $\tZ^\mu$. \eq{tZ} is a relatively simple equation that has
only one type of disconnectedness in the kernel, and unlike the corresponding
equation for the distinguishable particle case [see Eqs.(126) of I], \eq{tZ} is
not a matrix equation. We write \eq{tZ} as
\be
\tZ = G_0 - D_0tP_{12}\tZ              \eqn{tZ1}
\ee
where it is to be understood that $t=t_2$ and $D_0=d_3d_1$. Gauging this
equation gives
\be
\tZ^\mu=G_0^\mu-D_0^\mu t P_{12}\tZ-D_0t^\mu P_{12}\tZ - D_0tP_{12}\tZ^\mu
\ee
so that
\ben
(1+D_0tP_{12})\tZ^\mu=G_0^\mu-D_0^\mu tP_{12}\tZ-D_0 t^\mu P_{12}\tZ.
\een
With the help of \eq{tZ1} we then obtain
\ben
\tZ^\mu=\tZ G_0^{-1}G_0^\mu-\tZ G_0^{-1}D_0^\mu tP_{12}\tZ
-\tZ G_0^{-1}D_0 t^\mu P_{12}\tZ.
\een
Using
\ben
G_0^{-1}D_0=d^{-1}\hspace{5mm}\mbox{and}\hspace{5mm} G_0^\mu=d^\mu D_0+dD_0^\mu
\een
where $d=d_2$ and $d^\mu=d_2^\mu$, then gives
\ben
\tZ^\mu=\tZ d^{-1}d^\mu
+\tZ D_0^{-1}D_0^\mu G_0^{-1}\left(G_0-D_0tP_{12}\tZ\right)
-\tZ d^{-1}t^\mu P_{12}\tZ
\een
and therefore
\be
\tZ^\mu=\tZ d^{-1}d^\mu+\tZ\left( D_0^{-1}D_0^\mu D_0^{-1}d^{-1}
-d^{-1}t^\mu P_{12}\right) \tZ .                       \eqn{Zmu}
\ee
\begin{figure}[t]
\hspace*{1cm}  \epsfxsize=14cm\epsfbox{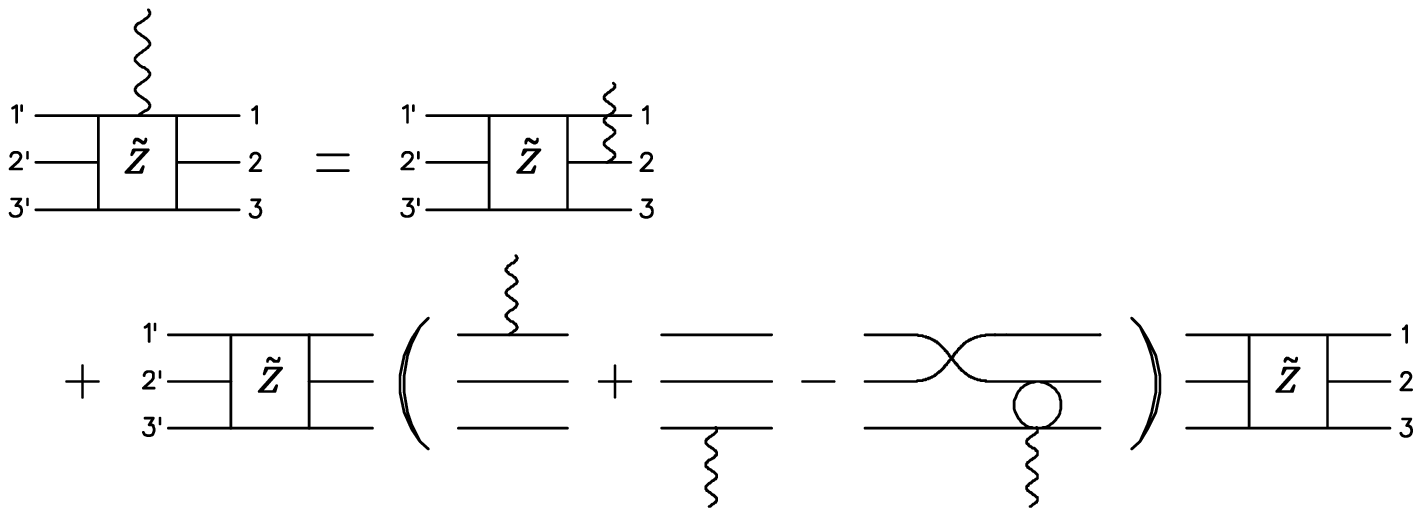}
\caption{\fign{Zgamma} Graphical representation of \protect\eq{Zmu} for the
gauged AGS-like Green function $\tZ^\mu$. }
\end{figure}
This equation, illustrated in \fig{Zgamma}, describes the attachment of photons
at all possible places in the multiple-scattering series of three identical
particles. As such, it forms the central result in the gauged identical
three-quark or three-hadron problem.  The structure of this equation may seem
surprising in that the second line is gauged only when it is right external [the
first term on the r.h.s. of \eq{Zmu}] whereas the first and third lines are
gauged everywhere (because $\tZ$ contains all possible diagrams, including
$G_0$).  That there is no inconsistency can be seen graphically from \fig{Z1}.
There we show a contribution to the term $\tZ\left( D_0^{-1}D_0^\mu
  D_0^{-1}d^{-1}\right)\tZ$ where the single scattering contribution to each
$\tZ$ is used and the intermediate line 1 is gauged. Because this is a Feynman
diagram, the gauging of the intermediate line 1 is the same as the gauging of
the left external line 2. To be noted is the crucial role the permutation
operator $P_{12}$ plays in the equation for $\tZ$, \eq{tZ} - this operator is
responsible for the crossing of the lines in the single scattering contributions
shown in \fig{Z1}.

\subsection{$Nd\rightarrow NNN$ transition current}

The $t$-matrix $T_{0d}$ for the breakup process $Nd\rightarrow NNN$ was given in
\eq{T0d_new}.  We follow the same procedure as given in Sec.\ III~C.4 for
distinguishable particles. Thus we do not gauge $T_{0d}$ directly but instead
introduce the Green function quantity
\be
\tT_{0d} = G_0 T_{0d} d_1 = \sum_{L_c}(P_{12}\tZ P_{12}-G_0) \phi_1.
\ee 
The electromagnetic current for $Nd\rightarrow \gamma NNN$ is then given by
\be
j_{0d}^\mu = G_0^{-1} \tT_{0d}^\mu d_1^{-1}.
\ee
Gauging $\tT_{0d}$, one obtains
\be
j_{0d}^\mu = \sum_{L_c}(P_{12}G_0^{-1}\tZ^\mu G_0^{-1} P_{12}
-G_0^{-1}G_0^\mu G_0^{-1})\psi_1
+\sum_{L_c}(P_{12} Z P_{12}-G_0^{-1})D_{01}\phi_1^\mu.
\ee
\begin{figure}[t]
\hspace*{5.5cm}  \epsfxsize=4.5cm\epsfbox{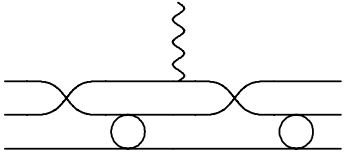}
\vspace{4mm}

\caption{\fign{Z1} A contribution to the term $\tZ\left( D_0^{-1}D_0^\mu
D_0^{-1}d^{-1}\right)\tZ$ showing how it leads to the gauging of the left
external particle 2.}
\end{figure}

\subsection{Three-nucleon bound-state current}

For three identical nucleons the electromagnetic bound-state current $j^\mu$ was
expressed in terms of the two-nucleon potential $v$ and gauged potential $v^\mu$
by \eq{howdy} and \eq{j^mu1_identical}. This was achieved by expressing the
seven-point function as $G^\mu=G\Gamma^\mu G$ and then taking the left- and
right-residues of $G^\mu$ at the three-body bound-state poles. The connection
with $j^\mu$ follows from the general structure of $G^\mu$ in the vicinity of
these poles:
\be 
G^\mu = \frac{i\Psi_K}{K^2-M^2} j^\mu \frac{i\bPsi_P}{P^2-M^2} . \eqn{Gjmu}
\ee
In this subsection we shall determine an alternative expression for $j^\mu$ that
is given in terms of the two-nucleon $t$-matrix $t$ and gauged $t$-matrix
$t^\mu$. To do this, we again take left- and right-residues of $G^\mu$ at the
three-body bound-state poles; however, this time we use an expression for
$G^\mu$ which is obtained by gauging \eq{Gid} where $G$ is written in terms of
the AGS Green function $\tU$. For the purpose of taking residues it is
sufficient to gauge just the connected part of $G$. Thus, making the choice
$\tU=-\tZ P_{12}$, we are led to the gauging of \eq{GUZ}:
\be
G_c^\mu = \sum_{L_cR_c} \left[ \left(G_0 T_1\right)^\mu\tZ T_2 G_0
+G_0 T_1\tZ^\mu T_2 G_0 + G_0 T_1\tZ \left(T_2 G_0\right)^\mu\right]. \eqn{Gc}
\ee
As discussed in Appendix C.2, $\tZ$ contains the three-nucleon bound-state pole
with the pole structure being given by \eq{Zpole}. The bound-state poles of
$G_c^\mu$ can therefore be revealed by simply using \eq{Zmu} to express the
middle term of \eq{Gc} (containing $\tZ^\mu$) in terms of $\tZ$ factors.  It is
seen that only this middle term contains both the left and right bound-state
poles; thus, in the vicinity of these poles one can write the seven-point
function as
\be 
G^\mu = \sum_{L_cR_c}G_0 T_1\frac{ i\Psi_2^K\bPsi_1^K}{K^2-M^2}\left(
D_0^{-1}D_0^\mu D_0^{-1}d^{-1} -d^{-1}t^\mu P_{12}\right)
\frac{i \Psi_2^P\bPsi_1^P}{P^2-M^2}T_2 G_0 .
\ee
Now using the results of Appendix B, we have that $\Psi_K = \sum_{L_c} G_0 T_1
\Psi_2^K$ and $\bPsi_P = \sum_{R_c}\bPsi_1^P T_2 G_0$. Comparing with \eq{Gjmu}
and using $\Psi_2=-P_{12}\bPsi_1$ then gives the desired expression
\be
\j^\mu = \bPsi_1^K \left[d_2^{-1}t_2^\mu -\left(\ggama_1^\mu d_3^{-1}
+\ggama_3^\mu d_1^{-1}\right) d_2^{-1}P_{12}\right]\Psi_1^P.
\eqn{jmu23}
\ee
Using the bound-state equations for $\Psi_1$ and
$\bPsi_1$ [\eq{Psi_i_idd} and \eq{bPsi_i_idd}], it is easy to see that
\be
\bPsi_1^K \ggama_1^\mu d_3^{-1}d_2^{-1}P_{12}\Psi_1^P
=\bPsi_1^K \ggama_2^\mu d_3^{-1}d_1^{-1}P_{12}\Psi_1^P.
\ee
We may thus write \eq{jmu23} as
\be
\j^\mu = \bPsi_1^K P_{12}d_1^{-1}t_1^\mu P_{12}\Psi_1^P
-\bPsi_1^K\left(\ggama_2^\mu d_3^{-1}d_1^{-1}
+\ggama_3^\mu d_1^{-1}d_2^{-1}\right)P_{12}\Psi_1^P  \eqn{j^mu_spec}
\ee
whose form corresponds to the bound-state current given by us in
Ref.\ \cite{nnn3d} in the context of the spectator approach to the three-nucleon
system.

\eq{j^mu_spec} may appear to be in a form where the first term on the right hand
side (RHS) corresponds to the two-body interaction current and the second term
corresponds to the one-body current. Yet this is not the case since $t^\mu$ in
fact contains both types of contribution. To see this explicitly we gauge \eq{t}
for $t$ in this way obtaining
\be
t^\mu = \frac{1}{2}tD_0^\mu t
+ \left(1+\frac{1}{2}tD_0\right)v^\mu\left(1+\frac{1}{2}D_0t\right).
\ee
The first term on the RHS corresponds to a one-body current while the second
term gives the two-body interaction current. Thus the total one-body current
contribution to the three-nucleon bound-state current is
\be
j^\mu_{\scriptsize\mbox{one-body}}
= \bPsi_1^K\left[P_{12}\frac{1}{2} d_1^{-1}t_1 D_0^\mu t_1 P_{12} -
\left(\ggama_2^\mu d_3^{-1}d_1^{-1}
+\ggama_3^\mu d_1^{-1}d_2^{-1}\right)P_{12}\right]\Psi_1^P.
\ee
Again using the bound-state equations for $\Psi_1$ and $\bPsi_1$, we may write
this result as
\be
j^\mu_{\scriptsize\mbox{one-body}}
= \bPsi_1^K \left(\ggama_2^\mu d_3^{-1}d_1^{-1}
+\ggama_3^\mu d_1^{-1}d_2^{-1}\right)\left(\frac{1}{2}-P_{12}\right)\Psi_1^P.
\ee

\subsection{Current conservation}

To prove current conservation for observables expressed in terms of $\tZ^\mu$,
it is useful to first deduce the WT identity for $\tZ^\mu$. To do this we write
\eq{Zmu} as
\be
\tZ^\mu=\tZ d_2^{-1}d_2^\mu+\tZ \Gamma_2^\mu \tZ       \eqn{Zmu_short}
\ee
where
\be
\Gamma_2^\mu = \left(\ggama_1^\mu d_3^{-1}
+\ggama_3^\mu d_1^{-1}\right) d_2^{-1}-d_2^{-1}t_2^\mu P_{12},
\ee
and then follow the procedure of I in Sec.\ III~B.2. Thus we introduce the
quantities
\be 
\he_i(k_1k_2k_3,p_1p_2p_3)=ie_i(2\pi)^{12}\delta^4(k_i-p_i-q)
\delta^4(k_j-p_j)\delta^4(k_k-p_k)
\ee
where $ijk$ represent a cyclic ordering of $123$. This allows us to write the WT
identities for the gauged two-body potential and gauged one-particle propagator
in three-particle space in terms of commutators as
\be
q_\mu v^\mu_i I_i=[\he_j+\he_k,v_i],\hspace{1cm}
q_\mu d^\mu_i I_j I_k=[\he_i,d_i]   \eqn{WT_comm}
\ee
where $I_i$, $I_j$, and $I_k$ are unit operators in the space of particles $i$,
$j$, and $k$ respectively. Using \eqs{WT_comm}, it is then easy to see that
\be
q_\mu t_2^\mu I_2 = [\he_3+\he_1,t_2],\hspace{1cm}
q_\mu \ggama_i I_j I_k = -[\he_i,d_i^{-1}].
\ee
The first term on the RHS of \eq{Zmu_short} contracted with $q_\mu$ thus gives
\be
q_\mu \tZ d_2^{-1}d_2^\mu = q_\mu \tZ \ggama_2^\mu d_2 =
-\tZ[\he_2,d_2^{-1}]d_2 = -\tZ[\he_2,G_0^{-1}]G_0,     \eqn{WT-Z1}
\ee
while for the last term of \eq{Zmu_short} we first deduce that
\be
q_\mu\Gamma_2^\mu = - [\he_3+\he_1,G_0^{-1}] - [\he_3+\he_1,t_2]d_2^{-1}P_{12}.
  \eqn{Gamma_3-wti}
\ee
By using
\be
\tZ^{-1}=G_0^{-1}+d_2^{-1}t_2P_{12}, \eqn{Z^-1}
\ee
which follows from \eq{tZ1}, the previous equation reduces to
\be
q_\mu\Gamma_2^\mu = - [\he_3+\he_1,\tZ^{-1}]
+ \left(G_0^{-1}-\tZ^{-1}\right)(\he_1-\he_2).
\ee
The WT identity for the last term of \eq{Zmu_short} thus becomes
\be
q_\mu \tZ\Gamma_2^\mu \tZ = [\he_3+\he_1,\tZ] - (\he_1+\he_2)\tZ +
\tZ G_0^{-1}(\he_1-\he_2)\tZ. \eqn{goose}
\ee
To simplify this expression further we use \eq{Z^-1} and the fact that
$[\he_2,\htt_2]=0$, to obtain
\bea
\tZ G_0^{-1}\he_1 \tZ &=& \he_1 \tZ - \tZ d_2^{-1} t_2 P_{12} \he_1 \tZ
= \he_1 \tZ - \tZ d_2^{-1} \he_2 t_2  P_{12}\tZ\nn
&=& \he_1 \tZ - \tZ G_0^{-1} \he_2 G_0 d_2^{-1} t_2  P_{12}\tZ\nn
&=& \he_1 \tZ + \tZ G_0^{-1} \he_2 \tZ - \tZ G_0^{-1} \he_2 G_0.
\eea
Substituting this result into \eq{goose}, we obtain
\be
q_\mu \tZ\Gamma_2^\mu \tZ = [\he,\tZ]+\tZ[\he_2,G_0^{-1}]G_0   \eqn{WT-Z2}
\ee
where $\he=\he_1+\he_2+\he_3$. Combining \eq{WT-Z2} with \eq{WT-Z1} we finally
obtain the WT for $\tZ^\mu$:
\be
q_\mu \tZ^\mu = [\he,\tZ].      \eqn{WT-Z}
\ee

The expression for the three-nucleon bound-state current $j^\mu$, \eq{jmu23},
can be written as
\be
j^\mu = \bPsi_1^K\Gamma_2^\mu\Psi_2^P
\ee
and on comparison with \eq{Zmu_short} is recognised to be the result of taking
simultaneous residues of $\tZ^\mu$ at the initial and final bound-state
poles. Taking such simultaneous residues of \eq{WT-Z}, the left hand side of
this equation gives $q_\mu j^\mu$, while the right hand side gives zero since
$\tZ$ has only a single pole; in this way we obtain the current conservation
equation for the bound-state current: $q_\mu j^\mu=0$.

To prove that the $Nd\rightarrow Nd$ electromagnetic transition current of
\eq{j_dd^mu} is conserved, we write the WT identity for $\phi_1^\mu$ and
$\bphi_1^\mu$ [see Eqs.(153) of I] in three-particle space as
\be
q_\mu \phi_1^\mu I_1 = (\he_2+\he_3)\phi_1, \hspace{1cm}
q_\mu \bphi_1^\mu I_1 = -\bphi_1(\he_2+\he_3),
\ee
and note that \eq{WT-Z} implies
\be
q_\mu\tU^\mu = [\he,\tU]  \eqn{WT-U}
\ee
where $\tU=-\tZ P_{12}$. Using the last three equations, we obtain from
\eq{j_dd^mu} that
\be
q_\mu j_{dd}^\mu =
d_1^{-1}\bphi_1[\he_1,\tU]\phi_1d_1^{-1}=
d_1^{-1}\he_1 d_1\bphi_1 d_2d_3 U d_2d_3\phi_1
-\bphi_1 d_2d_3 U d_2d_3\phi_1 d_1\he_1 d_1^{-1}.  \eqn{chook}
\ee
By shifting momentum arguments, the $\he_1$ factors stop the cancellation of
external $d_1^{-1}$ terms with the neighbouring $d_1$ propagators contained in
$\tU$.  Thus for on mass shell nucleons $d_1^{-1}=0$, and both terms on the RHS
of \eq{chook} become zero.

\subsection{Normalisation condition and charge conservation}

The normalisation condition for the three-body bound-state wave function that
was given in \eq{wave_norm_id} follows almost immediately from the basic
integral equation for $G$, \eq{Gas}. The expression obtained gives the
normalisation condition in terms of the three-body potential $V$ and therefore
in terms of input two-body potentials. The disadvantage of these quantities has
already been discussed. To obtain the normalisation condition without explicit
reference to $V$ we follow a similar procedure to that for $G$, but instead use
the AGS Green function $\tZ$. From \eq{Z^-1} it follows that
\be
\tZ = \tZ\left( G_0^{-1}+d_2^{-1}t_2 P_{12}\right)\tZ.
\ee
Using the pole structure of $\tZ$ given in \eq{Zpole} and then taking residues,
one obtains
\be
i\bPsi_1^P\frac{\partial}{\partial P^2}
\left( G_0^{-1}+d_2^{-1}t_2P_{12}\right)\Psi_2^P = 1
\ee
which is the normalisation condition expressed in terms of the input two-body
$t$-matrix and Faddeev components of the bound-state wave function.

Analogous to the normalisation condition for the three-body bound-state wave
function is a condition known as ``charge conservation'' for the three-body
bound-state electromagnetic current. The field theoretic definition of the
three-body bound-state current is given by \eq{j^mu_exp}. From this expression,
the translational invariance of the electromagnetic current operator, and the
fact that $|P\ra$ is an eigenstate of the charge operator, it is easy to show
that
\be j^\mu(P,P)= 2eP^\mu \eqn{charge}
\ee
where $e$ is the charge of the three-body bound state. \eq{charge} constitutes
the charge conservation condition and corresponds to the fact that the full
charge of the three particles is being probed by the external electromagnetic
probe. As such, it is an essential condition that needs to be satisfied by any
model calculation. Within our model the expression for $j^\mu$ is given by
\eq{j^mu_spec}. Assuming the Ward identities for the one- and two-body input, it
is a matter of simple algebra to prove that this expression does indeed satisfy
the charge conservation condition of \eq{charge}. Such a proof has already been
given by us in Ref.\ \cite{nnn3d} for the case of the three-nucleon bound-state
current within the spectator approach. As there is a direct one to one
correspondence between \eq{j^mu_spec} and the bound-state current in the
spectator approach [Eq.\ (25) of Ref.\ \cite{nnn3d}], this proof need not be
repeated here.

\section{SUMMARY}

We have applied the gauging of equations method, introduced in the preceding
paper for distinguishable particles \cite{I}, to the integral equations
describing the strong interactions of three identical relativistic particles.
For simplicity of presentation, we restricted the discussion to identical
particle systems like that of three quarks ($qqq$) or three nucleons (\NNN)
where there is no strong interaction coupling to two-particle channels.  Once
the strong interaction equations are specified, the gauging of equations method
couples an external photon to all possible places in the strong interaction
model, while at the same time preserving the proper identical particle symmetry
of the original equations. In this way we have obtained gauge invariant
expressions for the various electromagnetic transition currents of such
identical three-particle systems.

Two essentially different integral equations were gauged. The first was the
integral equation for the three-particle Green function $G$, \eq{G_as}. This
equation has a disconnected kernel that is defined in terms of the two-body
potential $v$.  As a result, the electromagnetic transition currents that follow
are themselves expressed in terms of $v$ and the gauged potential $v^\mu$.
Although formally correct, such expressions for the transition currents may not
be the most useful for practical calculations. We have therefore considered an
alternative formulation of the strong interaction problem in terms of
four-dimensional versions of the well known AGS equations. Such integral
equations have a kernel that is connected (after one iteration) and are
expressed in terms of two-body $t$-matrices. For identical particles there are
many ways to define the AGS amplitude, all giving the same three-body Green
function $G$. We have introduced an AGS amplitude that satisfies a particularly
simple AGS equation, \eq{tZ}, where the inhomogeneous term consists of a single
unpermuted term. This equation is ideal for our gauging procedure. By contrast,
the AGS equation used previously in three-nucleon calculations \cite{Glockle} is
inconvenient for gauging purposes as both its inhomogeneous term and kernel
consist of sums over two different permutations.  After gauging, our AGS
equation gives practical expressions in terms of the two-body $t$-matrix $t$ and
gauged $t$-matrix $t^\mu$, for all the electromagnetic transition currents of
three identical particles. In this sense, the gauging of equations method has
provided a unified description of the strong and electromagnetic interactions of
the identical three-particle system.

\acknowledgments
The authors would like to thank the Australian Research Council for their
financial support.

\appendix
\section{SYMMETRY PROPERTIES OF${\mbf{$\,\,\tilde{U}_{ij}$}}$ FOR
IDENTICAL PARTICLES}

Here we derive some useful properties of the AGS Green functions
$\tU_{ik}$ in the case of identical particles. Although these properties
are well known, we include them here both for completeness and to show how
they may be derived within the formalism used in this paper.
\bigskip

Using the definition of $\tU_{ik}$ given in \eq{G^D=U_ij}, it is easy to show
that
\be
\tU_{ik}=\bar{\delta}_{ik}G_0+\frac{1}{2}\sum_{j}\bar{\delta}_{ij}G_0V_j 
G_0\bar{\delta}_{jk}+\frac{1}{4}\sum_{jn}\bar{\delta}_{ij}G_0V_j 
G^DV_nG_0\bar{\delta}_{nk} .
\ee
For the case $k=1$ we explicitly have that
\bea
\tU_{11}&=&\frac{1}{2}G_0(V_2+V_3)G_0
+\frac{1}{4}G_0(V_2+V_3)G^D(V_2+V_3)G_0 \nn
\tU_{21}&=&G_0+\frac{1}{2}G_0V_3G_0
+\frac{1}{4}G_0(V_1+V_3)G^D(V_2+V_3)G_0 \eqn{U_i1} \\
\tU_{31}&=&G_0+\frac{1}{2}G_0V_2G_0
+\frac{1}{4}G_0(V_1+V_2)G^D(V_2+V_3)G_0 .\nonumber
\eea
From \eqs{V-ij-symm} and (\ref{G^D}) it follows that $G^D$ commutes with all
elements of the symmetry group $S_3$: $P_{ij}G^DP_{ij}=G^D$ (of course for $G$
there is the stronger symmetry property $P_{ij}G= GP_{ij}=-G$).
Now using the fact that
\be
P_{23}V_1P_{23}=V_1;\hspace{1cm} P_{23}V_2P_{23}=V_3;\hspace{1cm} 
P_{23}V_3P_{23}=V_2,
\ee
the following two relations are easily deduced from \eqs{U_i1}:
\be
P_{23}\tU_{21}P_{23}=\tU_{31};\hspace{1cm} P_{23}\tU_{31}P_{23}=\tU_{21}.
\ee
In a similar way one can deduce how any AGS Green function transforms under the
simultaneous exchange of two corresponding left- and right-particle labels. The
following is a complete list for the case $k=1$:
\bea
&&P_{12}\tU_{11}P_{12}=\tU_{22}\hspace{1cm} P_{12}\tU_{21}P_{12}=\tU_{12}
\hspace{1cm} P_{12}\tU_{31}P_{12}=\tU_{32}\nn
&&P_{23}\tU_{11}P_{23}=\tU_{11}\hspace{1cm} P_{23}\tU_{21}P_{23}=\tU_{31}
\hspace{1cm} P_{23}\tU_{31}P_{23}=\tU_{21}\nn
&&P_{31}\tU_{11}P_{31}=\tU_{33}\hspace{1cm} P_{31}\tU_{21}P_{31}=\tU_{23}
\hspace{1cm} P_{31}\tU_{31}P_{31}=\tU_{13}. \nonumber
\eea
The corresponding transformations for $k=2$ and $k=3$ can be written down by
inspection.

\section{THREE-BODY BOUND-STATE WAVE FUNCTION
FOR IDENTICAL PARTICLES}

The three-body Green function $G$ is defined by \eq{G6pt}. If the field theory
admits a three-body bound state, then $G$'s behaviour in the vicinity of the
bound-state pole is given by \eq{Gpole} where the bound-state wave function
$\Psi_P$ is defined in \eq{Psi_P}. These equations are true whether the three
particles are identical or not.

In the identical particle case the bound-state wave equation follows from
\eq{Gas}, $G=G_0^P+G_0VG$, by taking residues at the three-body bound-state
pole:
\be
\Psi = G_0V\Psi
\ee
where $V$ is given by \eq{V_sym}. In the absence of three-body forces,
\be
V=\frac{1}{2}(V_1+V_2+V_3)
\ee
and the Faddeev wave function components are defined by
\be
\Psi_i = \frac{1}{2}G_0 V_i\Psi    \eqn{Psi_i_id}
\ee
so that
\be
\Psi = \Psi_1 + \Psi_2 + \Psi_3 .     \eqn{Psi_sum}
\ee 
Using \eq{T_i} it then follows from \eq{Psi_i_id} that the $\Psi_i$ satisfy
the coupled equations
\be
\Psi_i = \frac{1}{2}G_0T_i\left(\Psi_j+\Psi_k\right) \eqn{Psi_i_coupled}
\ee
where $ijk$ are cyclic.

By writing \eq{Psi_i_id} explicitly showing particle labels, for example
\be
\Psi_1(123) = \frac{1}{2}D_0(23)v(23,2'3')\Psi(12'3'),
\ee
the following symmetry relations are easily deduced:
\bea
&&P_{23}\Psi_1 = -\Psi_1\hspace{2cm}
  P_{12}\Psi_1 = -\Psi_2\hspace{2cm}
  P_{31}\Psi_1 = -\Psi_3\nn
&&P_{31}\Psi_2 = -\Psi_2\hspace{2cm}
  P_{23}\Psi_2 = -\Psi_3\hspace{2cm}
  P_{12}\Psi_2 = -\Psi_1        \eqn{Psi_i_symm}   \\ 
&&P_{12}\Psi_3 = -\Psi_3\hspace{2cm}
  P_{31}\Psi_3 = -\Psi_1\hspace{2cm}
  P_{23}\Psi_3 = -\Psi_2.\nonumber
\eea
These symmetry relations enable us to write \eq{Psi_sum} as
\bea
\Psi(123) &=&  \Psi_1(123)+ \Psi_2(123)+ \Psi_3(123) \nn
&=&  \Psi_1(123)+ \Psi_1(231)+ \Psi_1(312),
\eea
or in general for any fixed $i$,
\be
\Psi = \sum_{P_c}  \Psi_i,
\ee
the sum being over cyclic permutations of the particle labels.
Similarly the symmetry relations \eqs{Psi_i_symm} enable us to write
\eqs{Psi_i_coupled} for $i=1$ as
\be
\Psi_1(123) = D_0(23)t(23,2'3')\Psi_1(2'3'1)
\ee
or in general for any $i$,
\be
\Psi_i = - G_0 T_i P_{ij} \Psi_i = - G_0 T_i P_{ik} \Psi_i  \eqn{Psi_i_idd}.
\ee
Similarly
\be
\bPsi_i = - \bPsi_i P_{ij} T_i G_0 = - \bPsi_i P_{ik} T_i G_0 \eqn{bPsi_i_idd}.
\ee

\section{THE GREEN FUNCTION $\tilde{\mbf{$\!Z$}}$}

In dealing with identical particles, only one AGS-like Green function $\tU$ is
needed to describe all possible processes of the three-body system. The
connected part of the three-body Green function, $G_c$, is expressed in terms of
$\tU$ according to \eq{GU_short}. In this appendix we present a derivation of
\eq{GU_short} and show how various forms for $\tU$ can be specified. One of
these involves the AGS-like Green function $\tZ$ which, because of its
simplicity, is chosen for the purposes of gauging. The pole structure of $\tZ$
is also discussed.

\subsection{Derivation}

Writing out the sums in \eq{GU_ik} and showing particle labels explicitly, this
equation can be written as
\bea
\lefteqn{G_c(123,1'2'3') = \frac{1}{4}
D_0(23)t(23,2''3'')\left[ \tU_{11}(12''3'',1'2'''3''')
t(2'''3''',2'3') D_0(2'3')\right.} \nn
&& + \left. \tU_{12}(12''3'',1'''2'3''')t(3'''1''',3'1') D_0(3'1')
 + \tU_{13}(12''3'',1'''2'''3')t(1'''2''',1'2') D_0(1'2')\right]P
\nn
&& + \frac{1}{4}D_0(31)t(31,3''1'')\left[ \tU_{21}(1''23'',1'2'''3''')
t(2'''3''',2'3') D_0(2'3')\right. \nn
&& + \left. \tU_{22}(1''23'',1'''2'3''')t(3'''1''',3'1') D_0(3'1')
 + \tU_{23}(1''23'',1'''2'''3')t(1'''2''',1'2') D_0(1'2')\right]P
\nn
&& + \frac{1}{4}D_0(12)t(12,1''2'')\left[ \tU_{31}(1''2''3,1'2'''3''')
t(2'''3''',2'3') D_0(2'3')\right. \nn
&& + \left. \tU_{32}(1''2''3,1'''2'3''')t(3'''1''',3'1') D_0(3'1')
 + \tU_{33}(1''2''3,1'''2'''3')t(1'''2''',1'2') D_0(1'2')\right]P
\nn
\eea
where integrals over double and triple dashed momenta are understood. The
symbols $P$ in the above equation indicate sums over all permutations of the
initial (right) particle labels. Carrying out these permutation sums and taking
into account the symmetries of the $\tU_{ij}$ discussed in Appendix A, this
equation can be simplified to read
\be
G_c(123,1'2'3') = \frac{1}{2} P_c \,D_0(23)t(23,2''3'')\tX(12''3'',1'2'''3''')
t(2'''3''',2'3')D_0(2'3')P_c       \eqn{GX}
\ee
where the symbols $P_c$ indicate sums over cyclic permutations of both initial
and final state particle labels, and where the Green function $\tX(123,1'2'3')$
is defined by
\be
\tX(123,1'2'3')=
\tU_{11}(123,1'2'3')+\tU_{12}(123,3'1'2')+\tU_{12}(132,2'1'3') .   \eqn{X_def}
\ee
Note the symmetry relation $\tX=P_{23}\tX P_{23}$.  The last two terms
of \eq{X_def} contribute equally to \eq{GX}; nevertheless, we do not define
$\tX$ as $\tU_{11}(123,1'2'3')+2\tU_{12}(123,3'1'2')$ as this would not make it
easy to write an integral equation for $\tX$.  Using the AGS equations for
$\tU_{11}$ and $\tU_{12}$, it follows that the $\tX$ of \eq{X_def} satisfies the
integral equation
\bea
\lefteqn{\tX(123,1'2'3') = G_0(123,3'1'2')+G_0(132,2'1'3')}\hspace{1cm}\nn
&+& \frac{1}{2}D_0(12)t(12,1''2'')\tX(31''2'',1'2'3')
+\, \frac{1}{2}D_0(31)t(31,3''1'')\tX(23''1'',1'2'3'). \eqn{X}
\eea
Writing \eq{GX} in the shorthand form as
\be
G_c = \frac{1}{2} P_c G_0 T_1 \tX T_1 G_0 P_c ,
\ee
we have in this way derived \eq{GU_short} with $\tU$ being specifically given by
$\tU=1/2\tX$. This form for $\tU$ has been used by Gl\"{o}ckle {\em et al.} in
practical calculations \cite{Glockle}.  However, it is possible to choose other
forms for $\tU$, and as we show below, one of these is particularly simple.

Since the two-body $t$-matrices in \eq{GX} are fully antisymmetric, we can
equally well write this equation as
\be
G_c(123,1'2'3') =\frac{1}{2} P_c\, D_0(23)t(23,2''3'')\tY(12''3'',1'2'''3''')
t(2'''3''',2'3')D_0(2'3')P_c       \eqn{GY}
\ee
where
\be
\tY(123,1'2'3') = \frac{1}{2}\left[ \tX(123,1'2'3')-\tX(132,1'2'3')\right].
\ee
We note that $\tY$ has the symmetry property $P_{23}\tY=\tY P_{23}=-\tY$
and satisfies the integral equation
\bea
\lefteqn{\tY(123,1'2'3') = 
\frac{1}{2}\left[ G_0(123,3'1'2')+G_0(123,2'3'1')
-G_0(132,3'1'2')-G_0(132,2'3'1')\right]}\hspace{1cm}\nn
&&+\, \frac{1}{2}D_0(12)t(12,1''2'')\tY(31''2'',1'2'3')
+ \frac{1}{2}D_0(31)t(31,3''1'')\tY(23''1'',1'2'3').
\eqn{Y}
\eea
Although this appears to be a more complicated equation than \eq{X}, it is
easily shown that $\tY$ can alternatively be specified as
\be
\tY(123,1'2'3') = \frac{1}{2} \left[ \tZ(123,3'1'2')
+\tZ(123,2'3'1')-\tZ(132,3'1'2')-\tZ(132,2'3'1')\right] \eqn{YZ}
\ee
where $\tZ$ satisfies the especially simple equation
\be
\tZ(123,1'2'3') = G_0(123,1'2'3') + D_0(31)t(31,3''1'')\tZ(23''1'',1'2'3').
\eqn{Z}
\ee
This equation constitutes a three-fold reduction in the size of the kernel in
comparison with the original AGS equations, \eqs{tildeU_sym}. 
One can write \eq{Z} in shorthand form as
\be
\tZ = G_0 - G_0  T_2 P_{12} \tZ = G_0 - G_0 P_{12} T_1  \tZ ,   \eqn{Z_sym1}
\ee
the last version of which is illustrated in \fig{Zfig}.
\begin{figure}[b]
\hspace*{2.0cm} \epsfxsize=13.0cm\epsfbox{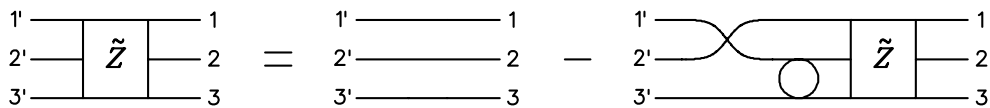}
\vspace{.4cm}

\caption{\fign{Zfig} Illustration of \protect{\eq{Z}} showing how the final
state momenta of the AGS-like Green function $\tilde{Z}$ are permuted in the
integral term.}
\end{figure}
It is recognised that the kernel of this equation is identical to the kernel
of the bound-state equation for $\Psi_2$. As \eq{Z_sym1} implies
$\tZ^{-1}=G_0^{-1}+T_2 P_{12}$,
it also follows that
\be
\tZ = G_0 - \tZ T_2 P_{12} G_0 = G_0 - \tZ P_{12} T_1  G_0 \eqn{Z_sym2}
\ee
which is an alternative equation for $\tZ$ whose kernel is identical to the
kernel of the bound-state equation for $\bPsi_1$.

Substituting \eq{YZ} into \eq{GY} and taking into account the antisymmetry of
the two-body $t$-matrices we obtain
\bea
\lefteqn{G_c(123,1'2'3') = \frac{1}{2}P_c\, D_0(23)t(23,2''3'')}\hspace{1cm}
\nn
&&\left[ \tZ(12''3'',3'''1'2''')+\tZ(12''3'',2'''3'''1')\right]
t(2'''3''',2'3')D_0(2'3')P_c .      \eqn{GUU}
\eea
The two $\tZ$ terms in \eq{GUU} in fact contribute equally to $G_c$.
This can be seen by using \eq{Z_sym2} to write
\bea
\tZ(123,3'1'2') &=& G_0(123,3'1'2')+\tZ(123,1''2''3')t(1''2'',1'2')D_0(1'2'),
\eqn{Z_1} \\
\tZ(123,2'3'1') &=& G_0(123,2'3'1')+\tZ(123,3''1''2')t(3''1'',3'1')D_0(3'1').
\eqn{Z_2}
\eea
When used in \eq{GUU}, the antisymmetry of the two-body $t$-matrices makes the
last terms of these equations identical, while the cyclic permutation operators
$P_c$ ensure the identity of their inhomogeneous terms. Using the antisymmetry
of $t$'s once more,  we arrive at our final form for the AGS Green function
$\tU$ that is to be used in \eq{GU_short}:
\be
\tU = -\tZ P_{12}
\ee
where $\tZ$ is given by \eq{Z}.  In view of the numerical simplicity of the
equation for $\tZ$, it is this form for $\tU$ that we choose in this paper
for the purposes of gauging.

\subsection{Pole structure of $\tilde{\mbf{$\!Z$}}$}

If the three-body system admits a three-body bound state, it is clear from
\eq{GU_short} that $\tU$, and therefore $\tZ$ has a pole at the bound-state
mass $M$. Writing the pole structure of $\tZ$ as
\be
\tZ(123,1'2'3') \sim i\frac{R(123,1'2'3')}{P^2-M^2},    \eqn{R}
\ee
it is our goal to deduce the explicit form of the residue $R(123,1'2'3')$.

Taking the residue of \eq{Z} at the bound-state pole gives the equation
\be
R(123,1'2'3') = D_0(31)t(31,3''1'')R(23''1'',1'2'3').
\ee
With the initial momenta fixed, this equation coincides with the equation for
$\Psi_2(123)$, the third Faddeev bound-state wave function component - 
see \eq{Psi_i_idd}. On the other hand, we can use \eq{Z_sym2} to write
\be
\tZ(123,1'2'3') = G_0(123,1'2'3')+\tZ(123,2''3''1')t(2''3'',2'3')D_0(2'3').
\ee
Taking residues of this equation gives
\be
R(123,1'2'3') = R(123,2''3''1')t(2''3'',2'3')D_0(2'3')
\ee
which for fixed final momenta coincides with the equation for $\bPsi_1(1'2'3')$
- see \eq{bPsi_i_idd}.  We may thus write $R(123,1'2'3') = C
\Psi_2(123)\bPsi_1(1'2'3')$ where $C$ is a constant. To determine $C$ we use
$\tU=-\tZ P_{12}$ in \eq{GU_short} and take residues of both sides of the
equation, in this way obtaining
\be
i \Psi\bPsi = - P_c G_0 T_1 C\Psi_2\bPsi_1 P_{12} T_1 G_0 P_c.
\ee
Using the results of Appendix B it is then easily seen that $C=i$. We have
thus shown that the pole structure of $\tZ$ is given by 
\be
\tZ(123,1'2'3') \sim i\frac{\Psi_2(123)\bPsi_1(1'2'3')}{P^2-M^2} .   \eqn{Zpole}
\ee


\begin{thebibliography}{99}
\bibitem{I} A.\ N.\ Kvinikhidze and B.\ Blankleider, nucl-th/98xxxxx.
\bibitem{gpinn} B.  Blankleider and A.  N.  Kvinikhidze, in Proceedings of the
APCTP Workshop on Astro-Hadron Physics, ``Properties of Hadrons in Matter'',
Seoul, Korea, 25-31 October, 1997, to be published by World Scientific,
nucl-th/9802017; a more detailed account is given in nucl-th/9810025, to be
published in Phys. Rev. C.
\bibitem{previous} A. N. Kvinikhidze and B. Blankleider, {\em Coupling photons
to hadronic processes}, invited talk at the Joint Japan Australia Workshop,
Quarks, Hadrons and Nuclei, November 15-24, 1995 (unpublished); in Proceedings
of the XVth International Conference on Few-Body Problems in Physics, Groningen,
The Netherlands, July 22-26, 1997, published in Nucl.\ Phys.\ {\bf A631}, 559c
(1998).
\bibitem{AT} See for example I.R. Afnan and A.W. Thomas in {\em Modern
Three-Hadron Physics}, Topics in Current Physics Vol. 2, edited by A.W. Thomas
(Springer-Verlag, Berlin 1977).
\bibitem{AGS} E. O. Alt, P. Grassberger, and W. Sandhas, Nucl. Phys. {\bf B2},
167 (1967).
\bibitem{KB4d} A.\ N.\ Kvinikhidze and B.\ Blankleider, Nucl.\ Phys.\ {\bf
A574}, 788 (1994).
\bibitem{WT} J. C. Ward, Phys. Rev. {\bf 78}, 182 (1950); Y. Takahashi,
Nuovo Cimento {\bf 6}, 371 (1957).
\bibitem{Glockle} W. Gl\"{o}ckle, {\em The Quantum Mechanical Few-Body Problem},
Texts and Monogr. Physics (Springer, Berlin, Heidelberg 1983); H. Wita{\l}a,
T. Cornelius and W. Gl\"{o}ckle, Few-Body Sys. {\bf 3}, 123 (1988);
D. H\"{u}ber, H. Wita{\l}a, and Gl\"{o}ckle, Few-Body Sys. {\bf 14}, 171 (1993).
\bibitem{nnn3d} A. N. Kvinikhidze and B. Blankleider, Phys. Rev. C {\bf 56},
2973 (1997).
\end{thebibliography}
\end{document}